\documentclass[aps,prd,groupedaddress,nofootinbib,amssymb,eqsecnum,showpacs,epsfig]{revtex4}
\usepackage{graphicx}
\usepackage{bm}
\usepackage{amsmath}

\def \lleq {\lower0.9ex\hbox{ $\buildrel < \over \sim$} ~}
\def \ggeq {\lower0.9ex\hbox{ $\buildrel > \over \sim$} ~}

\def \lcdm    {$\Lambda$CDM }
\def \lcdmb    {$\Lambda$CDM}

\def \om    {\Omega}

\def \omms   {\Omega_m}

\def \Geff {G_{\rm eff}}

\newcommand{\fs}{{\rm{\it f\sigma_8}}}
\newcommand{\magcir}{\raise
-3.truept\hbox{\rlap{\hbox{$\sim$}}\raise4.truept\hbox{$>$}\ }}
\newcommand{\mincir}{\raise
-3.truept\hbox{\rlap{\hbox{$\sim$}}\raise4.truept\hbox{$<$}\ }}

\def \beq  {\begin{equation}}
\def \eeq  {\end{equation}}
\def \ber  {\begin{eqnarray}}
\def \eer  {\end{eqnarray}}

\begin{document}
\newcommand{\newc}{\newcommand}

\newc{\be}{\begin{equation}}
\newc{\ee}{\end{equation}}
\newc{\ba}{\begin{eqnarray}}
\newc{\ea}{\end{eqnarray}}
\newc{\bea}{\begin{eqnarray*}}
\newc{\eea}{\end{eqnarray*}}
\newc{\D}{\partial}
\newc{\ie}{{\it i.e.} }
\newc{\eg}{{\it e.g.} }
\newc{\etc}{{\it etc.} }
\newc{\etal}{{\it et al.}}
\newcommand{\nn}{\nonumber}
\newc{\ra}{\rightarrow}
\newc{\lra}{\leftrightarrow}
\newc{\GB}{\mathcal{G}}
\newc{\lsim}{\buildrel{<}\over{\sim}}
\newc{\gsim}{\buildrel{>}\over{\sim}}
\title{Can the degeneracies in the gravity sector be broken?}

\author{Savvas Nesseris}
\email{savvas.nesseris@uam.es}

\affiliation{Instituto de F\'isica Te\'orica UAM-CSIC, Universidad Auton\'oma de Madrid,
Cantoblanco, 28049 Madrid, Spain}
\date{\today}

\begin{abstract}
It has been long known that by appropriately modifying gravity one can always reproduce the expansion history of any dark energy model, e.g. the $\Lambda$CDM. This degeneracy cannot be broken with geometric probes like the Type Ia supernovae, the cosmic microwave background or the baryon acoustic oscillations since they are based on the measurement of distances and scales and therefore require the expansion history $H(z)$, so one may only hope to break this degeneracy by using dynamic probes like the growth rate data that track the matter density perturbations. We demonstrate that breaking this degeneracy is not currently possible by explicitly constructing $f(R)$, $f(G)$ and $f(T)$ theories that mimic exactly the $\Lambda$CDM model at the background level and confronting them against the latest observational cosmological data. We also determine the necessary improvement in the growth rate data in order to discriminate these theories from $\Lambda$CDM. We found that at least a threefold improvement is necessary, something which will be possible with a survey like Euclid.
\end{abstract}


\maketitle

\section{Introduction}

Current observations suggest that the \lcdmb, a model based on general relativity (GR) that assumes a spatially flat geometry, a cosmological constant and cold dark matter, is consistent with nearly all cosmological observations. It is also based on the assumptions of homogeneity and isotropy on large cosmological scales and it has been implemented as an explanation of the accelerated expansion of the Universe. Most of the evidence in favor of the \lcdm comes from geometric tests that measure the expansion rate of the Universe $H(a)\equiv \dot{a}/a$ at different times, where $a(t)$ is the dimensionless scale factor of an Friedmann-Robertson-Walker metric. Examples of such geometric tests include measurements of the luminosity distance, $d_L(a)$, using standard candles like Type Ia supernovae (SnIa) \cite{Kilbinger:2008gk, Kessler:2009ys} and measurements of the angular diameter distance, $d_A(a)$, using standard rulers such as the last scattering horizon scale \cite{Komatsu:2010fb} and baryon acoustic oscillations \cite{Percival:2009xn}.

Despite the fact that these tests are currently the most accurate probes of dark energy, it is not possible to distinguish \lcdm from models that attribute the accelerating expansion to modifications of general relativity by just determining the Hubble parameter $H(a)$ \cite{Tsujikawa:2010zza, Heavens:2007ka, Nesseris:2006er,Nesseris:2006jc}. Furthermore, even if the background expansion is indeed exactly given by the \lcdmb, there is no \textit{a priori} guarantee that the sole cause for that is the cosmological constant $\Lambda$. This is due to the fact that once Pandora's box has been opened and the modifications of gravity have been ``released in the wild", one may always construct a \textit{degenerate} modified gravity model, e.g. based on the popular $f(R)$, $f(G)$ and $f(T)$ theories, that at the background level corresponds exactly to the \lcdm model, but at the perturbations level may be significantly different. For a similar discussion, but with interacting dark fluids, see Ref.~\cite{Aviles:2011ak}.

In order to address this problem, we need additional observational input which is not geometric in origin. One such probe is the growth function of the linear matter density contrast $\delta\equiv\frac{\delta\rho_m}{\rho_m}$. The main reason why these data are so helpful is the fact that the growth of matter density perturbations is mainly driven by the motion of matter and as a result is highly sensitive to both the expansion of the Universe $H(a)$ and any modifications of gravity beyond GR. Actually, it can be shown that the growth factor in modified gravity theories satisfies the following differential equation, on subhorizon scales ($k^2\gg a^2H^2$), where primes denote differentiation with respect to the scale factor $a$ \cite{DeFelice:2010gb, Tsujikawa:2007gd, DeFelice:2010aj, Nesseris:2009jf, Nesseris:2007pa}:
\be
\delta''(a)+\left(\frac{3}{a}+\frac{H'(a)}{H(a)}\right)\delta'(a)
-\frac{3}{2}\frac{\omms \Geff(a)/G_N}{a^5 H(a)^2/H_0^2}~\delta(a)=0.
\label{ode}
\ee
This differential equation has in general two solutions that correspond to two different modes, a growing and a decaying one, that in a matter dominated universe in GR behave as $\delta=a$ and as $\delta=a^{-3/2}$ respectively. In order to get the growing mode we demand that at early times $a_{in}$, usually during matter domination, the initial conditions have to be chosen as $\delta(a_{in})\simeq a_{in}$ and $\delta'(a_{in})\simeq1$. When $\Geff(a)/G_N=1$ we get GR as a subcase, while in general for modified gravity theories $\Geff$ can be time and scale dependent.

For a flat GR model with a constant dark energy equation of state $w$, the exact solution of Eq.~(\ref{ode}) for the growing mode is given by \cite{Silveira:1994yq, Percival:2005vm}
\ba
\delta(a)&=& a~{}_2F_1 \left(- \frac{1}{3 w},\frac{1}{2} -
\frac{1}{2 w};1 - \frac{5}{6 w};a^{-3 w}(1 - \omms^{-1})\right)
\label{Da1} \nn \\ \textrm{for}&&H(a)^2/H_0^2= \omms a^{-3}+(1-\omms)a^{-3(1+w)},\ea
where ${}_2F_1(a,b;c;z)$ is a hypergeometric function defined by the series
\be
{}_2F_1(a,b;c;z)\equiv \frac{\Gamma(c)}{\Gamma(a)\Gamma(b)}\sum^{\infty}_{n=0}\frac{\Gamma(a+n)\Gamma(b+n)}{\Gamma(c+n)n!}z^n \ee
on the disk $|z|<1$ and by analytic continuation elsewhere; see Ref.~ \cite{handbook} for more details. In more general cases it is impossible to find a closed form analytical solution for Eq.~(\ref{ode}), but in Ref.~\cite{Wang:1998gt} it was shown that the growth rate $f(a)\equiv \frac{d ln \delta}{dlna}$ can be approximated as
\ba
f(a)&=&\omms(a)^{\gamma(a)} \label{fg}\\
\omms(a)&\equiv&\frac{\omms~a^{-3}}{H(a)^2/H_0^2} \\
\gamma(a) &=&\frac{\ln f(a)}{\ln \omms(a)}\simeq \frac{3 (1-w)}{5-6 w}+\cdots .
\ea
We should note that the approximation for $\gamma$ is valid at first order for a dark energy model with a constant $w$, while for \lcdm ($w=-1$) we have $\gamma=\frac{6}{11}\simeq 0.545$.

The main goal of the present analysis is to test whether the current data can discriminate between these degenerate modified gravity models and a pure GR \lcdmb. Clearly, the most crucial role will be played by the growth rate data as they are the only ones that can probe the these theories at the perturbations level. Therefore, another interesting question that we will answer is what is the necessary improvement so that future growth rate data can irrefutably address this problem and break the degeneracy. In Sec. \ref{theory} we explicitly construct these models, which have only one free parameter $\alpha$, while in Sec. \ref{results} we present our main results and the forecasts for the potential of future growth rate data to differentiate between these theories and \lcdmb.

\section{Theory \label{theory}}
\subsection{The $f(R)$ model}
We will start from the action
\be
S= \frac{1}{8 \pi G} \int d^4 x \sqrt{-g}f(R)/2+S_m, \label{actionfR}
\ee
where $G$ is the bare gravitational constant, $f(R)$ is a function of the Ricci scalar  and $S_m$ is the action that corresponds to the matter Lagrangian $\mathcal{L}_m$. Varying with respect to the metric $g_{\mu\nu}$ we get the field equations
\be
\label{fieldeqfR}
F(R)R_{\mu \nu}-\frac{1}{2}f(R) g_{\mu \nu} -\nabla_\mu \nabla_\nu F(R)+g_{\mu \nu}\Box F(R)=8 \pi G T_{\mu \nu},
\ee
where $F_{R}\equiv \partial f/ \partial R$, $R_{\mu \nu}$ is the Ricci tensor and $T_{\mu \nu}$ is the energy momentum tensor for the matter fluid defined in Eq.~(\ref{en-mom-tensor}).

In a flat Friedmann-Robertson-Walker (FRW) metric with a scale
factor $a(t)$, we obtain the background (zero-order) equations:
\ba
\label{fRfried} 
3FH^2&=& (FR-f)/2-3H\dot{F}+8 \pi G \rho_m \\-2 F\dot{H}&=&\ddot{F}-H\dot{F}+8 \pi G (\rho_m+P_m),
\ea where $P_m$ and $\rho_m$ are the pressure and energy density of the matter fluid which satisfies the usual continuity equation
\be
\dot{\rho}+3 H (\rho+P)=0.
\ee
The \lcdm model satisfies
\be
H(a)^2=H_0^2\left(\omms a^{-3}+1-\omms\right), \label{HaGR}
\ee
while for the FRW metric we can express the Ricci scalar as
\be
R(a)=6(2 H(a)^2+a H(a)H'(a)).\label{Radef}
\ee
From these two equations we can express the Hubble parameter as
\be
H(R)^2=\frac{1}{3}\left(R-9(1-\omms)H_0^2\right)
\ee
and reexpress Eq.~(\ref{fRfried}) in terms of $R$
\be
f(R)+(R-6\Lambda) f'(R)-2(R-4 \Lambda)\left(1 + 3 (R - 3 \Lambda) f''(R)\right)=0,
\ee
where we have set $\Lambda=3 H_0^2(1-\omms)$. This differential equation describes all the Lagrangians $f(R)$ that have as a background the \lcdm model. The general solution can be found to be \cite{Multamaki:2005zs}, \cite{delaCruzDombriz:2006fj}
\begin{widetext}
\ba
f(R)&=&R-2\Lambda+C_1~\left(\frac{\Lambda }{R-3 \Lambda }\right)^{b_{1}} {}_2F_1\left(b_{1},\frac{3}{2}+b_{1},\frac{13}{6}+2b_{1},\frac{\Lambda }{R-3 \Lambda }\right)\nn\\&+&C_2\left(\frac{\Lambda }{R-3 \Lambda }\right)^{b_{2}} {}_2F_1\left(b_{2},\frac{3}{2}+b_{2},\frac{13}{6}+2b_{2},\frac{\Lambda }{R-3 \Lambda }\right),\label{fRmodact}
\ea
\end{widetext} where $b_{1/2}=\frac{1}{12} \left(-7\mp\sqrt{73}\right)$, $C_1$ and $C_2$ are constants to be determined from the data. In Eq.~(\ref{fRmodact}), the terms $R-2\Lambda$ correspond to the GR action plus a cosmological constant. Obviously, since for this model the expansion history satisfies Eq.~(\ref{HaGR}) by definition, no geometric test can differentiate it from GR. The only way to do that is to use dynamic tests, like the growth rate of perturbations. In $f(R)$ theories we have that \cite{Tsujikawa:2007gd}
\ba
\Geff / G_N&=&\frac{1}{F}\frac{1+4\frac{k^2}{a^2}m}{1+3\frac{k^2}{a^2}m}, \label{GefffR}\\ m&\equiv& \frac{F_{,R}}{F},\\F&\equiv&f_{,R}=\frac{\partial f}{\partial R},
\ea
which reduces to GR only when $f(R)=R-2\Lambda$. We will not show here the exact form of $\Geff$ that corresponds to the Lagrangian of Eq.~(\ref{fRmodact}) as it is rather complicated, but it can be obtained trivially with simple differentiations. We should note that Eq. (\ref{ode}) with $\Geff$ given by (\ref{GefffR}) is strictly valid only under the subhorizon approximation and that an improved version without so strong assumptions was presented in Ref.~\cite{delaCruzDombriz:2008cp}. However, for viable $f(R)$ models that are close to \lcdm or are exactly the same at the background level like in our case, Eq.~(\ref{ode}) is a very good approximation indeed \cite{DeFelice:2010gb}.

We should note that if we want to recover GR at early times ($a\ll1$), i.e. $\Geff / G_N\sim 1$ or $f'(R)\sim 1$, then we need to set $C_1=0$. With this choice of the parameters then it is easy to see that the Lagrangian of Eq.~(\ref{fRmodact}) passes all criteria for viability found in Ref. \cite{Pogosian:2007sw}. Then our Lagrangian is,
\begin{widetext}
\be
f(R)=R-2\Lambda+\alpha~H_0^2\left(\frac{\Lambda }{R-3 \Lambda }\right)^{b_{2}} {}_2F_1\left(b_{2},\frac{3}{2}+b_{2},\frac{13}{6}+2b_{2},\frac{\Lambda }{R-3 \Lambda }\right),\label{fRmodact1}
\ee
\end{widetext}
where now the parameter $\alpha$ is dimensionless and will be determined by the data in the next sections. Also, it should be noted that this Lagrangian was also studied in \cite{Dunsby:2010wg}, where it was argued that both the parameters $C_1$ and $C_2$  have to be set to zero in order to have a real valued $f(R)$. However, it is easy to see that this is not the case as from Eqs. (\ref{HaGR}) and (\ref{Radef}) we have that $R\geq4\Lambda$ and as a result, the last term in Eq.~(\ref{fRmodact1}) is always real.

Finally, it should be stressed that this Lagrangian is more than a mathematical curiosity as it is in principle viable, since it passes all the criteria of \cite{Pogosian:2007sw}, and it is really degenerate at the background with respect to \lcdmb, as can easily be checked numerically. Also, it is not an artifact of the construction method presented above as it has a clear contribution at the perturbations level since $\Geff$ is evolving and therefore by using cosmological data one should be able to discriminate it from the usual \lcdmb.

\subsection{The $f(\mathcal{G})$ model \label{fG}}
We will start from the action
\be
S= \frac{1}{8 \pi G_N} \int d^4 x \sqrt{-g}\left(R/2+f(\mathcal{G})\right)+S_m, \label{action}
\ee
where $G_N$ is the bare gravitational constant, $f(\GB)$ is a function of the Gauss-Bonnet term $\GB=R^2-4 R_{\mu\nu}R^{\mu\nu}+R_{\mu\nu \sigma\rho}R^{\mu\nu \sigma\rho}$ and $S_m$ is the action that corresponds to the matter Lagrangian $\mathcal{L}_m$. Varying with respect to the metric $g_{\mu\nu}$ we get the field equations \cite{DeFelice:2009rw}
\begin{widetext}
\be
\label{fieldeq}
G_{\mu \nu}+8\left[ R_{\mu \rho \nu \sigma} +R_{\rho \nu} g_{\sigma \mu}
-R_{\rho \sigma} g_{\nu \mu} -R_{\mu \nu} g_{\sigma \rho} +R_{\mu \sigma} g_{\nu \rho}+(R/2)(g_{\mu \nu} g_{\sigma \rho}-g_{\mu \sigma} g_{\nu \rho})\right] \nabla^{\rho} \nabla^{\sigma} f_{\GB}+(\GB f_{\GB}-f) g_{\mu \nu}=8\pi G_N\; T_{\mu \nu},
\ee
\end{widetext}
where $f_{\GB}\equiv \partial f/\partial \GB$, $G_{\mu \nu}$ is the Einstein tensor and $T_{\mu \nu}$ is the energy momentum tensor for the matter fluid defined as
\be T_{\mu\nu}=-\frac{2}{\sqrt{-g}}\frac{\delta(\sqrt{-g}
\mathcal{L}_m)}{\delta g^{\mu\nu}}, \label{en-mom-tensor}
\ee
In what follows we will only consider a matter fluid with an equation of state $w=0$, i.e. pressureless dust.

In a flat FRW metric with a scale factor $a(t)$, we obtain the background (zero-order) equations:
\ba
\label{mogfried} 3H^2&=&\GB f_{\GB}-f-24H^3 \dot{f_{\GB}}+8\pi G_N \rho_m \nn\\&=& \GB f_{\GB}-f-24H^3 f_{\GB \GB}\dot{\GB}+8\pi G_N \rho_m,
\ea
where $f_{\GB \GB}\equiv \partial^2 f/\partial \GB^2$, $H\equiv \frac{\dot{a}}{a}$ and $\rho_m$ is the energy density of the matter fluid which satisfies the usual continuity equation
\be \dot{\rho}+3 H \rho=0.
\ee
In this setup the Gauss-Bonnet term is given by
\ba
\GB&=&24H^2 \frac{\ddot{a}}{a}\nn\\
&=&24H^2(H^2+\dot{H}) \nn\\
&=&24H^2(H^2+a H H'(a)).
\label{GB}
\ea

Now, we will describe the procedure for the construction of a Lagrangian that at the background level gives an evolution exactly equal to the $\Lambda$CDM model. To achieve this we will start from the modified Friedman equation and demand that the extra terms due to the Gauss-Bonnet terms correspond to the contribution of $\Lambda$, i.e. they are constant, and then solve a differential equation to find to which Lagrangian $f(G)$ it corresponds. Specifically, we start from Eq.~(\ref{mogfried}) and rewrite it in the form:
\ba
\label{mogfried1} H(a)^2/H_0^2&=&\frac{1}{3} H_0^{-2}(\GB f_{\GB}-f-24H^3 f_{\GB\GB}\dot{\GB})+\om_m a^{-3} \nn\\&=& \om_{GB}(a) +\om_m a^{-3}.
\ea
where
\ba
\om_{GB}(a) &\equiv& \frac{ H_0^{-2}}{3}\left(\GB f_{\GB}-f-24H^3 f_{\GB\GB}\dot{\GB}\right)\nn\\
&=&\frac{ H_0^{-2}}{3}\left(\GB f_{\GB}-f-24 a H^4 f_{\GB\GB}\GB '(a)\right).~~
\ea 
Then we can demand that the term $\om_{GB}(a)$ corresponds to the contribution from  $\Lambda$, i.e. $\om_{GB}(a)=1-\om_m$, or
\be
\GB f_{\GB}-f-24a H^4 f_{\GB\GB}\GB '(a)-3(1-\om_m)H_0^2=0. \label{mog1}
\ee
Remembering that $G'(a)=1/a'(G)$ and writing everything in terms of $\GB$ we get
\be
\GB f_{\GB}-f(G)-24\frac{a(G)}{a'(G)}H(G)^4 f_{\GB\GB}-3(1-\om_m)H_0^2=0. \label{mog2}
\ee
In order to solve the previous equation we try the following ansatz:
\be
f(\GB)=-3H_0^2 (1-\om_m)+\beta_1\;\GB+\beta_2~\GB \int \frac{h(\GB)}{\GB^2} d\GB/ \label{myaction}
\ee
Substituting Eq.~(\ref{myaction}) to (\ref{mog2}) we get an equation for $h(\GB)$ as
\be
h'(\GB)=\frac{\GB \; a'(\GB)}{24 a(\GB) H(\GB)^4}\;h(\GB),
\ee
which can easily be solved to give
\ba
h(\GB)=h_0\;e^{\int_{\GB_0}^\GB \frac{\GB a'(\GB)}{24 a(\GB) H(\GB)^4} \, d\GB},\label{solhG1}
\ea
where $\GB_0=\GB(a=1)$ and $h_0=h(\GB_0)$. Replacing the stand-alone term $\GB$ in Eq.~(\ref{solhG1}) with its definition (\ref{GB}), we can then see that the part inside the integral is equal to $\frac{a'(\GB)}{a(\GB)}+\frac{H'(\GB)}{H(\GB)}$ and the integral can easily be done to yield
\be
h(\GB)=h_0\;a(\GB) H(\GB)/H_0.
\ee
Then the Lagrangian of Eq.~(\ref{myaction}) becomes
\be
f(\GB)=-3H_0^2 (1-\om_m)+\alpha~H_0^2\GB \int \frac{a(\GB) H(\GB)/H_0}{\GB^2} d\GB , \label{myaction1}
\ee
where the first term corresponds to the cosmological constant, we have neglected the term that was just proportional to $\GB$ as it does not contribute in the field equations, and finally we note that the second term corresponds to the  new contribution. Clearly, due to its construction this term does not contribute at the background level but only at the perturbations level where, as we will see in the next section, it exhibits interesting phenomenology. Also, it is easy to check with direct substitution in the modified Friedmann equation (\ref{mogfried}) that the solution with the Lagrangian given by (\ref{myaction1}) indeed is the $\Lambda$CDM. Finally, this model only has one free parameter $\alpha$, which is dimensionless, besides $\om_m$.

The cosmological perturbations of this class of theories were studied in Ref.~\cite{DeFelice:2009rw}. We can define the matter density perturbation $\delta_m$ as \be \delta_m \equiv  \frac{\delta \rho_m}{\rho_m}+\frac{\dot{\rho}_m}{\rho_m} \upsilon \label{drm}\ee where $\rho_m$ is the background matter density, $\delta \rho_m$ is the first-order perturbation of the density and $\upsilon$ is a velocity potential. In Ref.~\cite{DeFelice:2009rw} it was shown that $\delta_m$ satisfies the evolution equation (using the subhorizon approximation $k\gg aH$):

\be
\label{per1}
\ddot{\delta}_m+C_1(k,a) \dot{\delta}_m+C_2(k,a) \delta_m\approx 0,
\ee
where the functions $C_1(k,a)$ and $C_2(k,a)$ where first derived in \cite{DeFelice:2009rw} and are given in Appendix \ref{pert} for completeness. In the GR limit Eq.~(\ref{per1}) reduces to \be \ddot{\delta}_m+2 H \dot{\delta}_m-\frac{3}{2} \om_m a^{-3} \delta_m= 0 \ee so comparing these two expressions we can define an effective  Newton's constant: 
\be 
G_{eff}(k,a)/G_N= \frac{C_2(k,a)}{-\frac{3}{2} \om_m a^{-3}}, \label{geff}
\ee 
which is valid only under the subhorizon approximation $k\gg aH$.

In general, modified gravity models suffer from instabilities and ghosts due to the presence of the higher derivative terms. Many authors have studied these for the $f(R,\GB)$ class of theories, see for example Ref.~\cite{Quiros:2010ei}; however the most relevant instabilities for our discussion are the ones in the matter density perturbations during the matter era. In \cite{DeFelice:2009rw} is was shown that if one defines the deviation parameters
\ba
\xi &\equiv& f_{,\GB},\\
\mu &\equiv& H\dot{\xi}, \label{mueq1}
\ea
where $\mu$ is dimensionless and $\xi$ is constant only for the $\Lambda$CDM model, then the negative instabilities appear at \cite{DeFelice:2009rw}
\be
\mu \approx (aH/k)^2,\label{mueq2}
\ee
while for possibly viable models the maximum value of the deviation parameter is $\mu_{max}\lesssim10^{-6}$. Of course, as it was mentioned in \cite{DeFelice:2009rw} the UV limit of this class of theories is still unsatisfactory and they are considered ruled out in the literature, but we still study them in the context of the present analysis as simple toy models. In our case, for the Lagrangian of Eq.~(\ref{myaction1}) the deviation parameter is $\mu=\frac{\alpha a}{24 H(a)/H_0}$, which implies the quite strong constraint $\alpha \lesssim \frac{H(a)/H_0}{a}\;24\cdot10^{-6}$.

\subsection{The $f(T)$ model \label{fT}}
We now proceed to the case of the $f(T)$ gravity, where the
action can be written as \cite{Zheng:2010am}, \cite{Linder:2010py} (and references therein)
\begin{eqnarray}
\label{action11}
 I = \frac{1}{16\pi G_N }\int d^4x e\left[T+f(T)\right]+S_m,
\end{eqnarray}
where the $S_m$ corresponds to the contribution of matter
with energy densities $\rho_m$.

Now, assuming a universe governed by $f(T)$ gravity, that also is homogeneous and isotropic,
we can make the common choice for the vierbiens, that is
\begin{equation}
\label{weproudlyuse}
e_{\mu}^A={\rm
diag}(1,a,a,a),
\end{equation}
which corresponds to a flat FRW background geometry with metric
\begin{equation}
ds^2= dt^2-a^2(t)\,\delta_{ij} dx^i dx^j,
\end{equation}
with $a(t)$ the scale factor. Varying the action (\ref{action11}) with
respect to the vierbeins we acquire the field equations

\be\label{eom}
e^{-1}\partial_{\mu}(ee_A^{\rho}S_{\rho}{}^{\mu\nu})[1+f_{T}] +
e_A^{\rho}S_{\rho}{}^{\mu\nu}\partial_{\mu}({T})f_{TT}-[1+f_{T}]e_{A}^{\lambda}T^{\rho}{}_{\mu\lambda}S_{\rho}{}^{\nu\mu}+\frac{1}{4} e_ { A} ^ {\nu}[T+f({T})] = 4\pi Ge_{A}^{\rho}\overset {\mathbf{em}}T_{\rho}{}^{\nu},
\ee
where $f_{T}=\partial f/\partial T$, $f_{TT}=\partial^{2} f/\partial T^{2}$,
and $\overset{\mathbf{em}}{T}_{\rho}{}^{\nu}$  stands for the usual
energy-momentum tensor. Inserting the vierbein choice (\ref{weproudlyuse}) into the field equations (\ref{eom}) we obtain the modified Friedmann equations
\ba
\label{background1}
&&H^2= \frac{8\pi G_N}{3}\rho_m-\frac{f}{6}+\frac{Tf_T}{3},\\\label{background2}
&&\dot{H}=-\frac{4\pi G_N(\rho_m+P_m)}{1+f_{T}+2Tf_{TT}},
\ea
where
$H\equiv\dot{a}/a$ is the Hubble parameter, with dots denoting
derivatives with respect to the cosmic time $t$. We mention that in order to
bring the Friedmann equations closer to their standard form, we
 used   the relation
\begin{eqnarray}
\label{TH2}
T=-6H^2,
\end{eqnarray}
which arises straightforwardly for an FRW universe.

Comparing the first Friedmann equation (\ref{background1}) to the usual one,
we can see that we can obtain an effective dark energy sector of (modified)
gravitational origin. In particular, one can define the
dark energy density and pressure as \cite{Linder:2010py}:
\begin{eqnarray}
&&\rho_{DE}\equiv\frac{3}{8\pi
G_N}\left[-\frac{f}{6}+\frac{Tf_T}{3}\right], \label{rhoDDE}\\
\label{pDE}
&&P_{DE}\equiv\frac{1}{16\pi G_N}\left[\frac{f-f_{T} T
+2T^2f_{TT}}{1+f_{T}+2Tf_{TT}}\right],
\end{eqnarray}
while the effective equation-of-state parameter reads
\begin{eqnarray}
\label{wfT}
 w
=-\frac{f/T-f_{T}+2Tf_{TT}}{\left[1+f_{T}+2Tf_{TT}\right]\left[f/T-2f_{T}
\right] }.
\end{eqnarray}
Also, we can define
\begin{eqnarray}
\label{TH3}
E^{2}(z)\equiv\frac{H^2(z)} {H^2_{0}}=\frac{T(z)}{T_{0}},
\end{eqnarray}
where $T_0\equiv-6H_{0}^{2}$.

Unlike $f(R)$ gravity, the effective Newton's constant
in $f(T)$ gravity is not affected by the scale $k$, but rather surprisingly it
takes the following simple form \cite{Zheng:2010am}
\begin{eqnarray}
\label{GeffT}
\frac{G_{\rm eff}(a)}{G_{N}}=\frac{1}{1+f_{T}}.
\end{eqnarray}

In order to construct an $f(T)$ Lagrangian that at the background level gives a cosmic evolution exactly equal to that of the \lcdm model, we can start from the modified Friedmann equation (\ref{background1}) and demand that the last two terms are constant and equal to $\;H_0^2\left(1-\om_m\right)$, which is the contribution made by $\Lambda$, or
\ba
-\frac{f}{6}+\frac{T f_T}{3}=H_0^2\left(1-\om_m\right).
\ea
We can easily solve this equation to get \cite{Zheng:2010am}
\ba
f(T)=-2\Lambda +C_1 \sqrt{-T}=-2\Lambda+\alpha \;H_0 \sqrt{-T},\label{fTLag}
\ea
where $\alpha=\frac{C_1}{H_0}$ is a dimensionless parameter to be determined from the data in the next sections. Clearly the first term is the usual cosmological constant, while the second does not contribute at the background level, but it does contribute at the perturbations level as from Eqs.~(\ref{GeffT}) and (\ref{fTLag}) we have
\ba
\frac{\Geff(a)}{G_{N}}&=&\frac{1}{1+f_{T}}=\frac{1}{1-\frac{\alpha}{2 \sqrt{6} \; H(a)/H_0}}\nn\\
&=&\frac{1}{1-\frac{\alpha}{2 \sqrt{6}\;E(a)}},
\label{GeffT1}
\ea
where $E(z)$ is given by Eq.~(\ref{TH3}). As it can be seen, $\Geff(a)$ has the following three limiting values: at early times or $a\ll1$ we have $\Geff/G_N\rightarrow1$, today ($a=1$) we have $\Geff/G_N=\frac{1}{1-\frac{\alpha }{2 \sqrt{6}}}$ and at the distant future $a\gg1$ we have $\Geff/G_N\sim \frac{1}{1-\frac{\alpha }{2 \sqrt{6} \sqrt{1-\omms}}}$.

Also, in this case it is possible to find an analytical solution to Eq.~(\ref{ode}) when the expansion history is given by the \lcdm model
\be
E(z)^2\equiv H(z)^2/H_0^2=\omms a^{-3}+1-\omms, \label{Ezlcdm}
\ee
We provide the details for the derivation in Appendix \ref{analsol} and only show the final result for the growth factor $\delta(z)$ here
\begin{widetext}
\be
\delta(z)= K_1 \delta_1(z) +K_2 \delta_2(z), \label{fTanalsoldz}
\ee
\end{widetext}
where we have defined the functions $\delta_1$-$\delta_2$ as
\ba
\delta_1(z)&=& E(z)-\tilde{\alpha},\\
\delta_2(z)/\delta_3(z)&=& -4 \tilde{\alpha}  \sqrt[3]{E(z)-\tilde{\alpha} } \sqrt[3]{E(z)^2-\delta \Omega}\; F_1\left(\frac{2}{3};\frac{1}{3},\frac{1}{3};\frac{5}{3};\frac{\tilde{\alpha} +\sqrt{\delta \Omega }}{\tilde{\alpha} -E(z)},\frac{\sqrt{\delta \Omega }-\tilde{\alpha} }{E(z)-\tilde{\alpha} }\right)\nn\\&-&4 \delta \Omega +2^{2/3} (E(z)-\tilde{\alpha} ) \left(\sqrt{\delta \Omega }-E(z)\right) \sqrt[3]{\frac{E(z)}{\sqrt{\delta \Omega }}+1} \; _2F_1\left(\frac{1}{3},\frac{2}{3};\frac{5}{3};\frac{1}{2}-\frac{E(z)}{2 \sqrt{\delta \Omega }}\right)+4 E(z)^2,\\
\delta_3(z)&=&\frac{1}{4 \left(\delta \Omega -\tilde{\alpha} ^2\right) \sqrt[3]{E(z)^2-\delta \Omega }},
\ea
and we have also set $\tilde{\alpha}\equiv\frac{\alpha}{2\sqrt{6}}$, $\delta \Omega\equiv 1-\omms$, $E(z)$ is given by Eq.~(\ref{Ezlcdm}), $F_1(a;b_1,b_2;c,x,y)$ is the Appell hypergeometric function of two variables and ${}_2F_1(a,b;c;z)$ is the usual hypergeometric function.\footnote{The Appell hypergeometric function of two variables $F_1(a;b_1,b_2;c,x,y)$ can be implemented in Mathematica as AppellF1$[a,b_1,b_2,c,x,y]$ while the function ${}_2F_1(a,b;c;z)$ can be implemented as Hypergeometric2F1$[a,b,c,z]$. Both functions can be evaluated to arbitrary precision for real and complex arguments.} In order to get the appropriate behavior for the growing mode, i.e. $\delta\sim a$ for $a\ll1$, the constants $K_1$ and $K_2$ have to be defined as
\ba
K_1&=& \frac{K_2\;\Gamma \left(-\frac{1}{6}\right) \Gamma \left(\frac{5}{3}\right) \sqrt[6]{\delta \Omega }}{4 \sqrt{\pi } \left(\delta \Omega -\tilde{\alpha} ^2\right)},\label{K1}\\
K_2&=& -\frac{5}{3} \Omega_m^{1/3},\label{K2}
\ea
where $\Gamma(z)=\int_0^\infty t^{z-1}e^{-t} dt$ is the usual Euler gamma function.
At this point, it should be stressed again that this solution is only valid for the $f(T)$ model when $\Geff$ is given by Eq.~(\ref{GeffT1}) and the cosmic expansion is given by the \lcdm model of Eq.~(\ref{Ezlcdm}). Finally, we have checked that there is perfect agreement between this analytical solution and a numeric one.

\section{Results\label{results}}
\subsection{Comparison with the data\label{resultscomp}}
In this section we will compare the different theories presented in the previous sections against the data and see if we can discriminate them against GR and \lcdmb. We will use the same data [SnIa, cosmic microwave background (CMB), BAO, and growth rate] as in Refs. \cite{Basilakos:2013nfa} and \cite{Nesseris:2013jea}. For details in the analysis see the aforementioned references; however, there is an important difference in the fitting of the growth rate data. Instead of using various \textit{ad hoc} parametrizations for $\gamma(z)$ and then modeling the growth rate as $f(z)=\Omega(z)^{\gamma(z)}$, we instead fit the numerical solution of Eq.~(\ref{ode}) directly. Also, we should mention a few more things about the growth rate data, given in Table I of Ref. \cite{Basilakos:2013nfa}.

The growth data used in the present analysis are based on the {\em WiggleZ}, SDSS, 2dF, PSCz, VVDS, 6dF, 2MASS and BOSS galaxy surveys. The data points themselves are given in terms of the combination $f(z)\sigma_{8}(z)$, where $f(z)\equiv \frac{d\ln\delta(a)}{d\ln a}$ is the growth rate of structure\footnote{The growth rate $f(z)$ should not be confused with the functions $f(R)$ or $f(T)$ that correspond to the Lagrangians of the modified gravity models.} and $\sigma_8(z)$ is the redshift-dependent rms fluctuations of the linear density field. The advantage of using $f\sigma_{8}\equiv \fs$, instead of just $f(z)$, is that it is an almost model-independent way of expressing the observed growth history of the Universe (see \cite{Song09}). It should also be noted that the observed growth rate of structure ($f_{obs}=\beta~b$) can be calculated from the redshift space distortion parameter $\beta(z)$ and the linear bias $b$.

From the observational point of view, it is possible to use the anisotropy of the correlation function to estimate the $\beta(z)$ parameter; however, the linear bias can also be expressed as the ratio of the variances of the tracer (galaxies, QSOs, etc) and underlying mass density fields, smoothed at $8h^{-1}$ Mpc $b(z)=\sigma_{8,tr}(z)/\sigma_{8}(z)$, where $\sigma_{8,tr}(z)$ is measured directly from the sample. Using these definitions we can arrive at $\fs\equiv f \sigma_{8}=\beta \sigma_{8,tr}$. It should be mentioned that the different cosmologies, including modified gravity, affect the observational determination of $\beta(z)$ (and thus of $\fs$) very weakly, and only through the definition of distances.

Specifically, for the \lcdm we found $\chi^2_{min}=575.291$ for $\omms=0.272\pm0.003$, while for the $f(R)$ model of Eq.~(\ref{fRmodact1}) we found $\chi^2_{min}=574.347$ for $(\omms,\alpha)=(0.272\pm0.003,0.008 \pm 0.013)$ and for the $f(T)$ model of Eq.~(\ref{fTLag}) we found $\chi^2_{min}=573.885$ for $(\omms,\alpha)=(0.272\pm0.003,0.555\pm0.461)$. Clearly, in all cases the best-fit value of $\omms$ is dominated by the geometric probes and in particular the CMB.

We will also consider the Akaike information criterion (AIC) \cite{Akaike1974} as in Refs. \cite{Basilakos:2013nfa} and \cite{Nesseris:2013jea}. The AIC is defined,
for the case of Gaussian errors, as:
\be
{\rm AIC}=\chi^2_{min}+2k
\ee
where $k$ is the number of free parameters. A smaller value of the AIC indicates a better fit to the data; however, small differences in AIC are not necessarily significant, so in order to effectively compare the different models, we have to estimate the differences $\Delta {\rm AIC} = {\rm AIC}_{1} - {\rm AIC}_{2}$, for the two models $1$ and $2$. The higher the value of $|\Delta{\rm AIC}|$, the higher the evidence against the model with higher value of ${\rm AIC}$, with a difference $|\Delta {\rm AIC}| \magcir 2$ indicating a positive such evidence and $|\Delta {\rm AIC}| \magcir 6$ indicating a strong such evidence, while a value $\mincir 2$ indicates consistency among the two comparison models. In an nutshell, the AIC penalizes models with more parameters. We found the following values for the differences $\Delta {\rm AIC}={\rm AIC}_i-{\rm AIC}_\Lambda=(0,1.056,0.594)$ for the $\Lambda$CDM, the $f(R)$ and $f(T)$ models respectively. So, statistically speaking, all three models are at an equal footing with the current data.

In the case of the $f(G)$ model we found that the bestfit was practically indistinguishable from that of the \lcdm model and that the parameter $\alpha$ was constrained to be $\alpha\sim0.0045$. For values outside the region $|\alpha| \lesssim 0.005$ the differential equation (\ref{ode}) suffered from the instabilities mentioned in the previous section; as for $|\alpha| \gtrsim 0.005$ we have $\mu\gtrsim0.00005$, which is much larger than the constraint $\mu_{max}\lesssim 10^{-6}$. Therefore, the allowed parameter space is almost nonexistent and as result in what follows we will only focus in the other two cases, the $f(R)$ and $f(T)$ models.

\begin{figure*}[t!]
\centering
\vspace{0cm}\rotatebox{0}{\vspace{0cm}\hspace{0cm}\resizebox{0.48\textwidth}{!}{\includegraphics{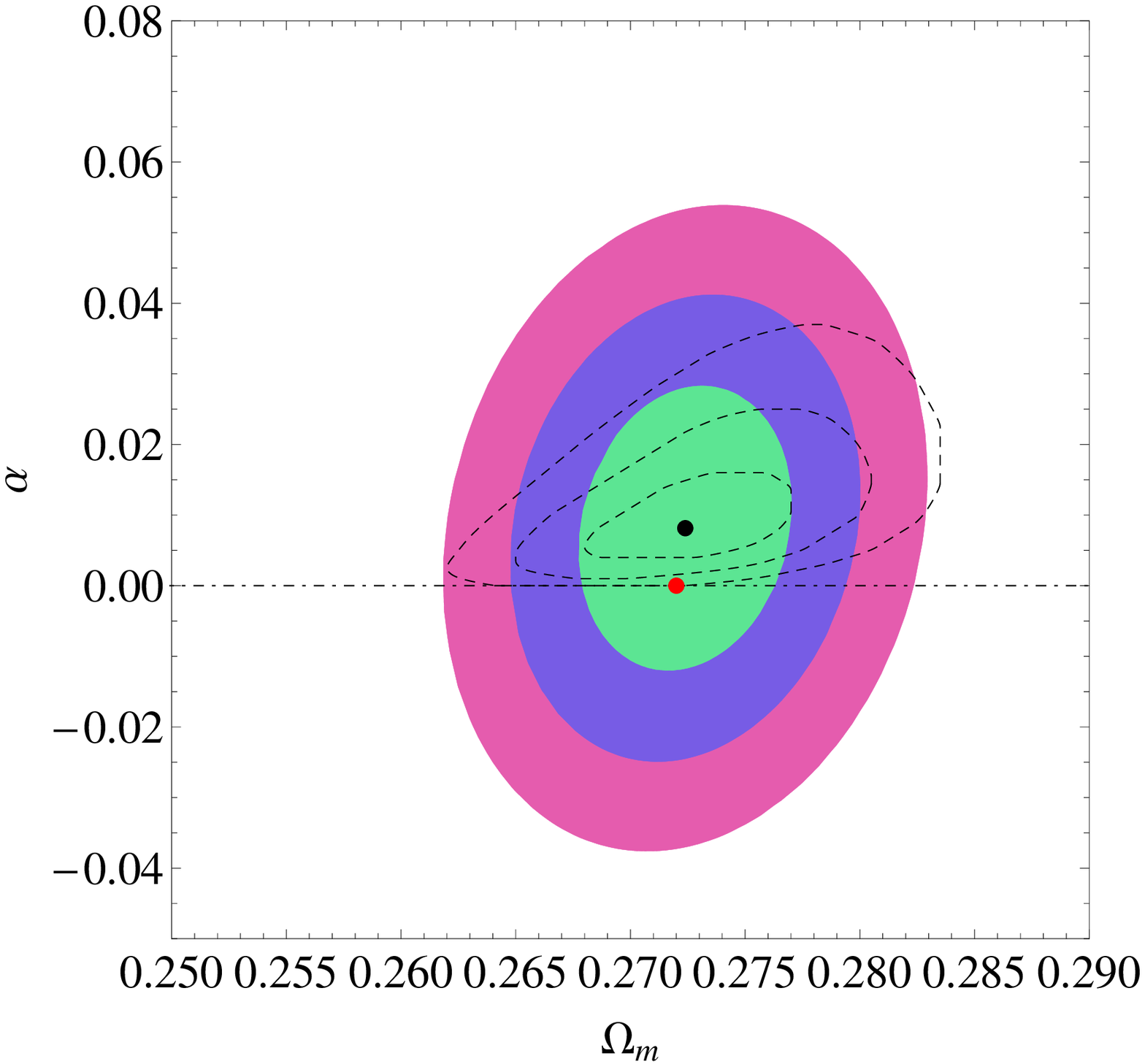}}}
\vspace{0cm}\rotatebox{0}{\vspace{0cm}\hspace{0cm}\resizebox{0.455\textwidth}{!}{\includegraphics{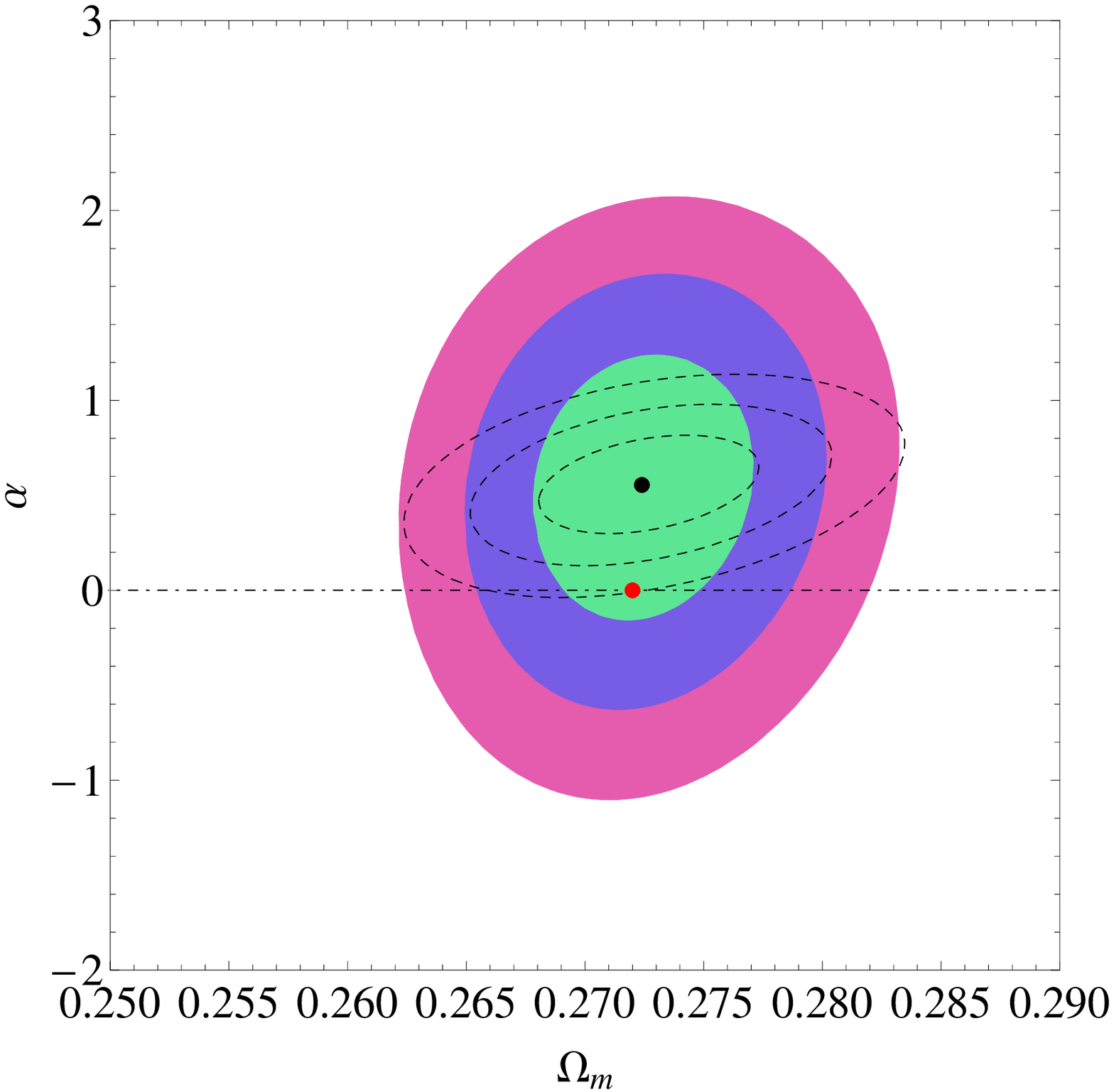}}}
\caption{Likelihood contours at the 1$\sigma$, 2$\sigma$ and
$3\sigma$ confidence levels, in the $(\Omega_{m},\alpha)$ plane for the $f(R)$ (left) and $f(T)$ (right) models. The black dot corresponds to the best fit $f(R)$ and $f(T)$ models, while the red dot to the \lcdm best-fit. The dashed contours indicate the case such that $\alpha=0$ sits exactly at the edge of the $3\sigma$ contour due to an improvement in the future growth rate data by a factor $s$, signifying a $3\sigma$ difference between these models and GR. \label{contours}}
\end{figure*}

In Fig. \ref{contours} we show the likelihood contours at the 1$\sigma$, 2$\sigma$ and
$3\sigma$ confidence levels in the $(\Omega_{m},\alpha)$ plane for the $f(R)$ (left) and $f(T)$ (right) models. The black dot corresponds to the $f(R)$ and $f(T)$ best fit, while the red dot to the \lcdm best fit. The dashed contours indicate the case such that $\alpha=0$ sits exactly at the edge of the $3\sigma$ contour due to an improvement in the future growth rate data by a factor $s$, signifying a $3\sigma$ difference between these models and GR, something we will elaborate more upon in the next section.

\begin{figure*}[t!]
\centering
\vspace{0cm}\rotatebox{0}{\vspace{0cm}\hspace{0cm}\resizebox{0.48\textwidth}{!}{\includegraphics{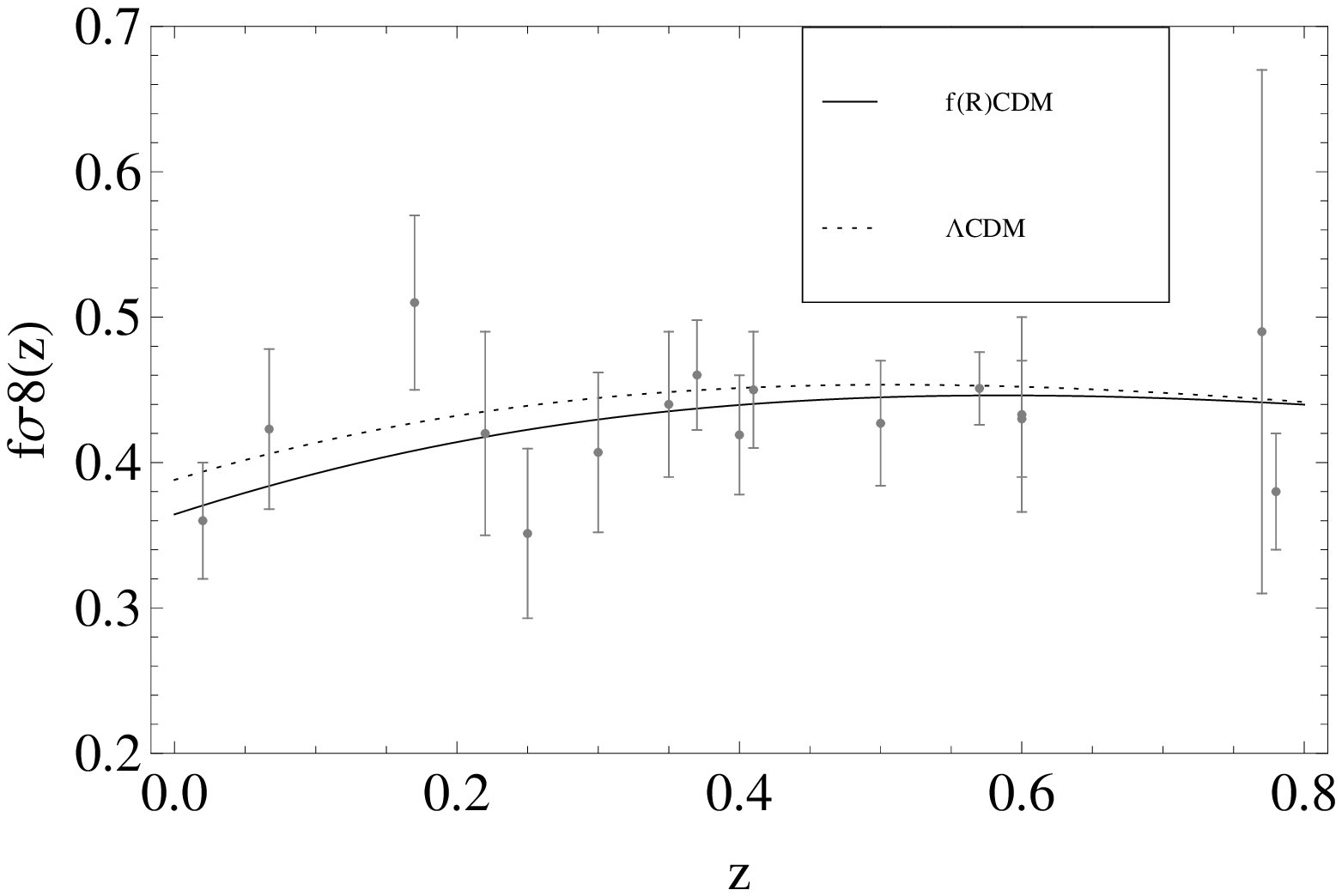}}}
\vspace{0cm}\rotatebox{0}{\vspace{0cm}\hspace{0cm}\resizebox{0.48\textwidth}{!}{\includegraphics{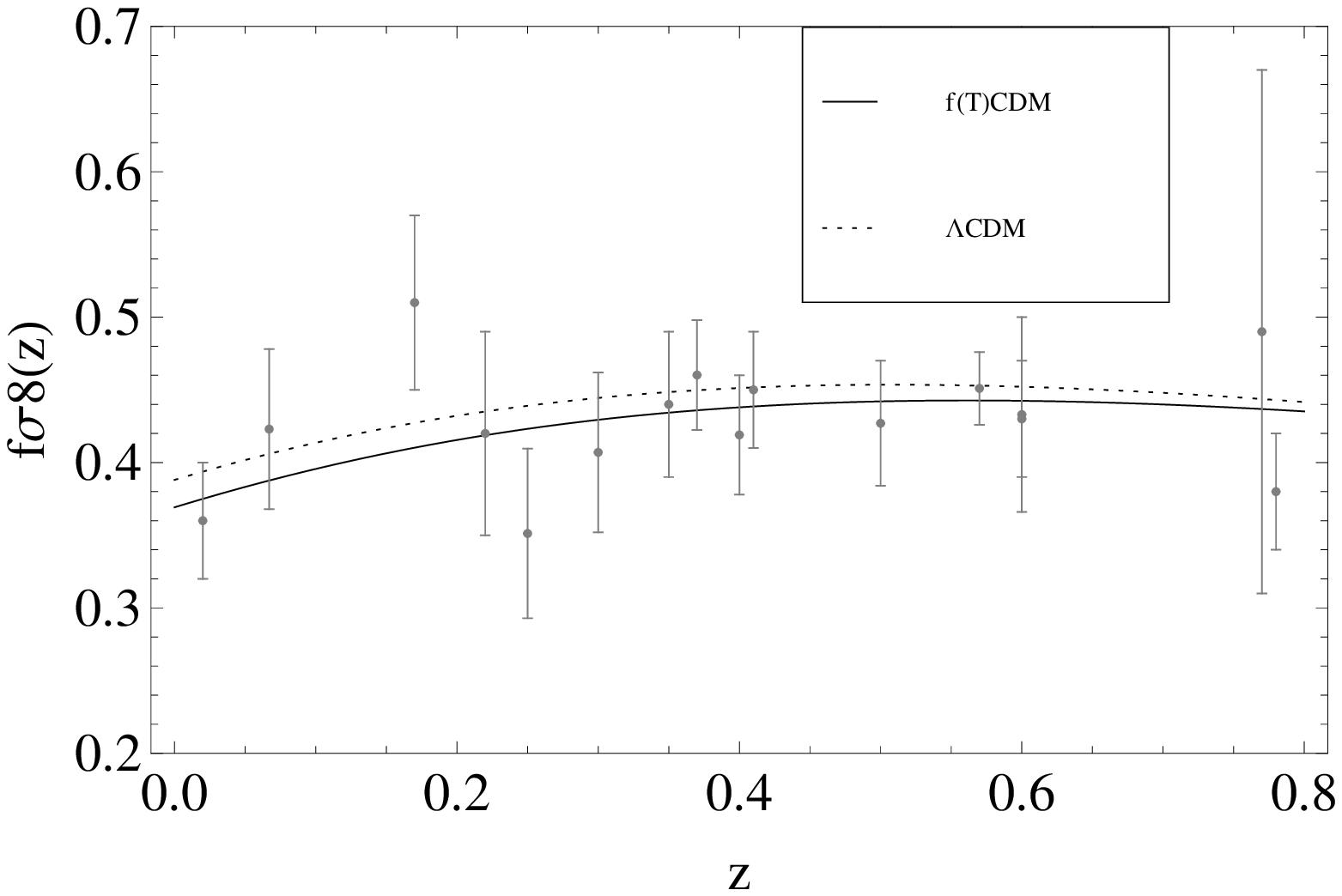}}}
\caption{Comparison of the observed and theoretical evolution of the growth rate $f\sigma_8(z)=F(z)\sigma_{8}(z)$ for the $f(R)$ (left) and $f(T)$ (right) models. The solid black lines correspond to the best-fit $f(R)$ and $f(T)$ models, while the dotted black lines correspond to the exact solution of Eq.~(\ref{ode}) for $f\cdot\sigma_8(z)$ for the $\Lambda$CDM model for $\Omega_{m}=0.272$.\label{growthrate}}
\end{figure*}

\begin{figure*}[t!]
\centering
\vspace{0cm}\rotatebox{0}{\vspace{0cm}\hspace{0cm}\resizebox{0.48\textwidth}{!}{\includegraphics{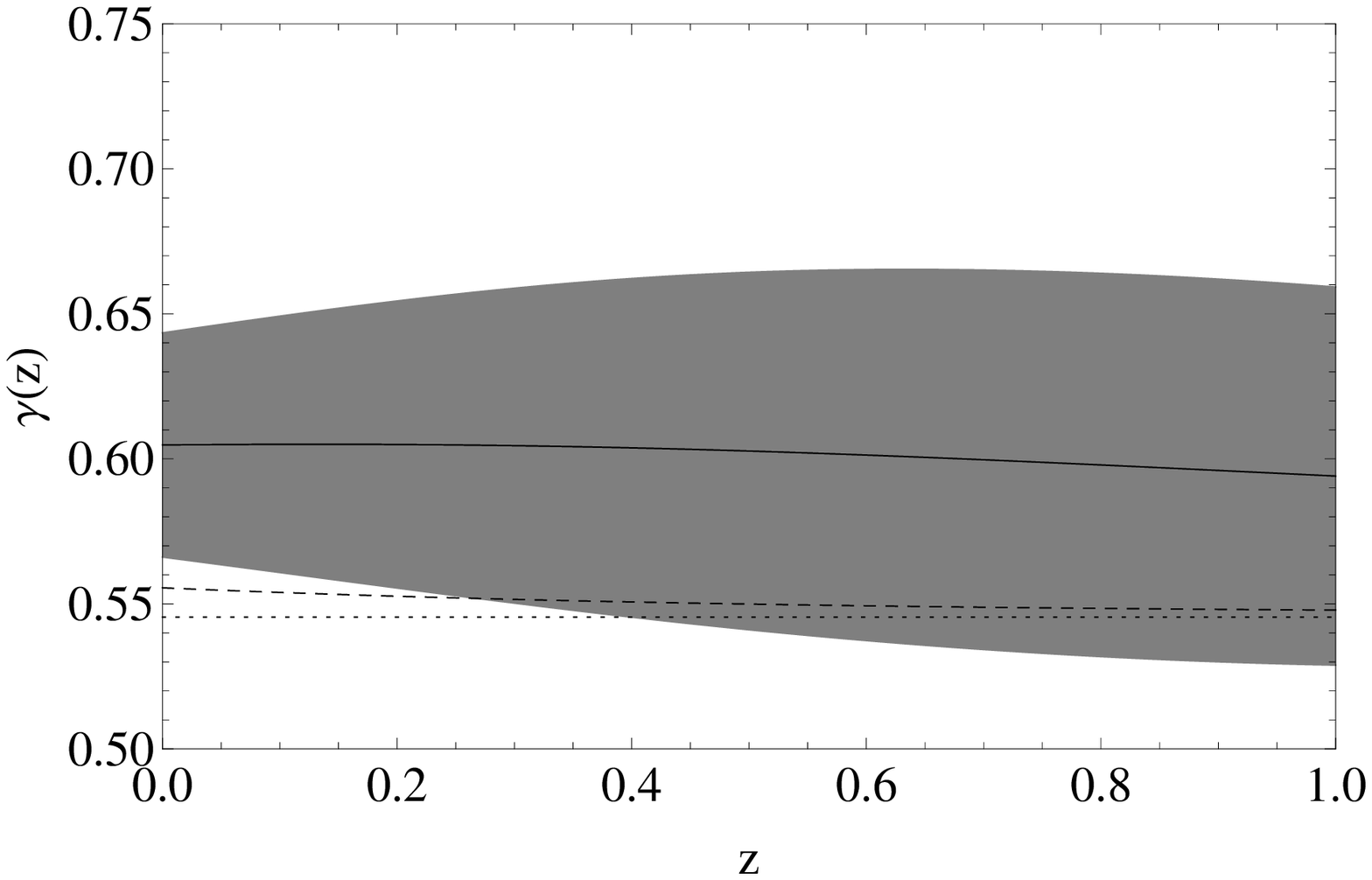}}}
\vspace{0cm}\rotatebox{0}{\vspace{0cm}\hspace{0cm}\resizebox{0.48\textwidth}{!}{\includegraphics{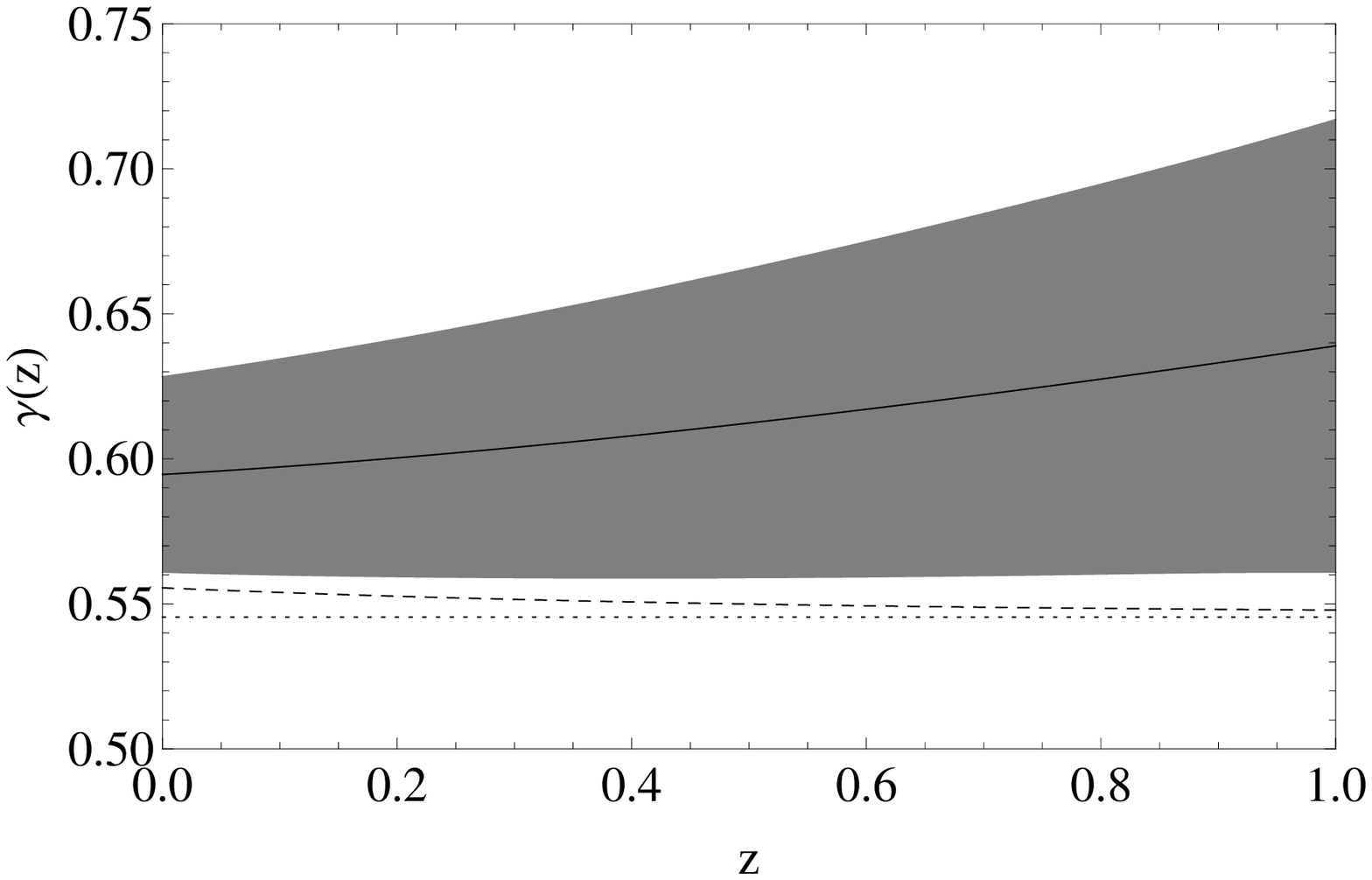}}}
\caption{Comparison of $\gamma(z)$ for the $f(R)$ (left) and $f(T)$ (right) models. The dotted line corresponds to the theoretical value $\gamma=6/11$ for \lcdmb, while the dashed line to its exact numerical value. The solid black line corresponds to the best fit and the gray regions to the $1\sigma$ error region.  \label{gammaplot}}
\end{figure*}

In Fig. \ref{growthrate} we show a comparison of the observed and theoretical evolution of the growth rate $f\sigma_8(z)=F(z)\sigma_{8}(z)$ for the $f(R)$ (left) and $f(T)$ (right) models. The dotted lines correspond to the best-fit $f(R)$ and $f(T)$ models, while the solid black lines correspond to the exact solution of Eq.~(\ref{ode}) for $f \sigma_8(z)$ for the $\Lambda$CDM model for $\Omega_{m}=0.272$. Note that at higher redshifts both models tend asymptotically to the \lcdm model, while at low redshifts there is significant difference. Related to this, in Fig.~\ref{gammaplot} we show the parameter $\gamma(z)=\frac{\ln f(z)}{\ln \Omega(z)}$ for the $f(R)$ (left) and $f(T)$ (right) models as a function of the redshift $z$. The dotted line corresponds to the theoretical value $\gamma=6/11$ for \lcdmb, while the dashed line corresponds to its exact numerical value. The solid black line corresponds to the best fit and the gray regions to the $1\sigma$ error region. In this case we found that the value of the parameter $\gamma_0\equiv\gamma(z=0)$ is compatible to the \lcdm model at the $1\sigma$ level. Specifically, for the f(R) model we found $\gamma_0=0.605\pm0.039$, while for the $f(T)$ model we found $\gamma_0=0.595\pm0.034$.

\begin{figure*}[t!]
\centering
\vspace{0cm}\rotatebox{0}{\vspace{0cm}\hspace{0cm}\resizebox{0.48\textwidth}{!}{\includegraphics{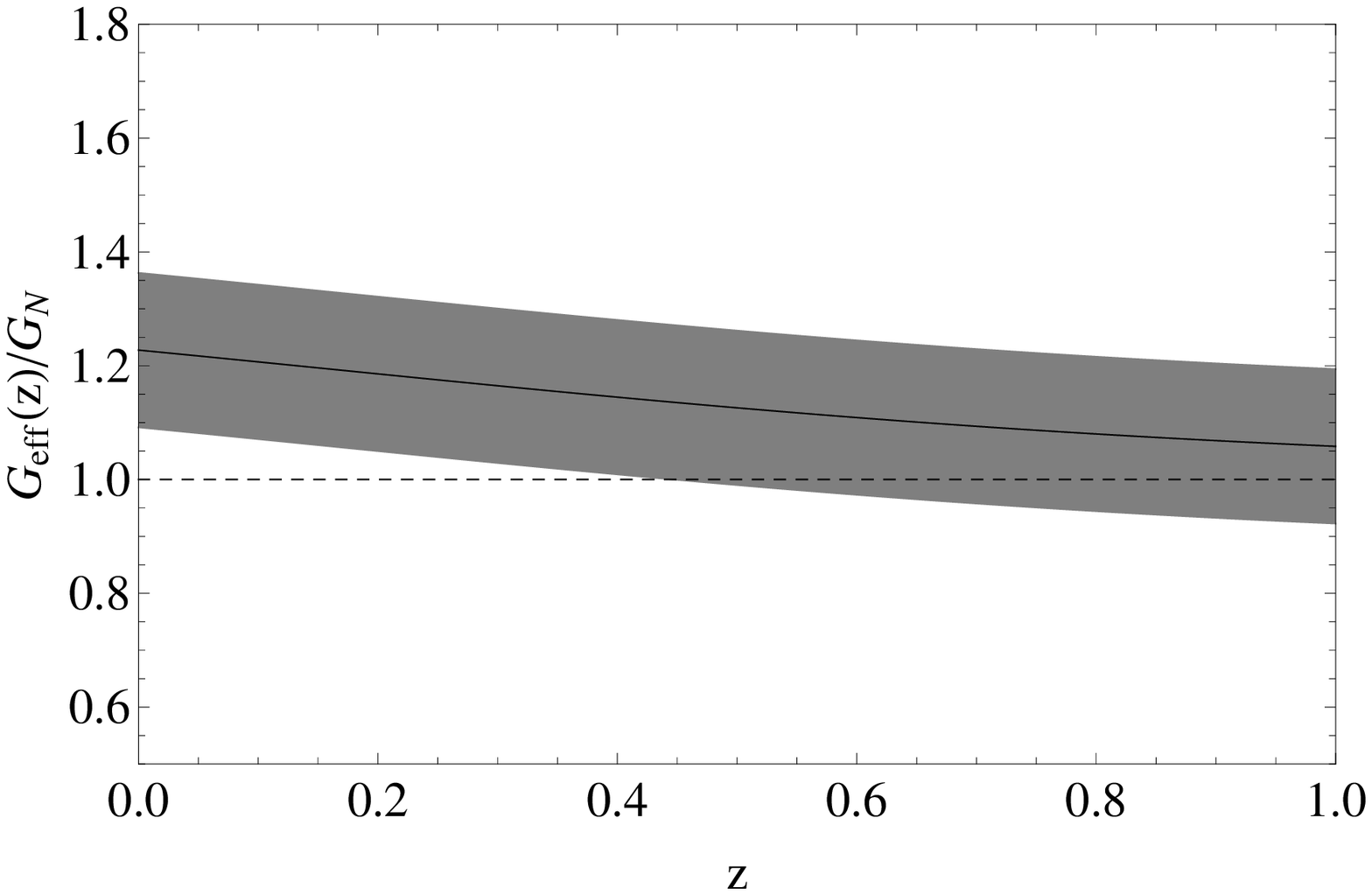}}}
\vspace{0cm}\rotatebox{0}{\vspace{0cm}\hspace{0cm}\resizebox{0.48\textwidth}{!}{\includegraphics{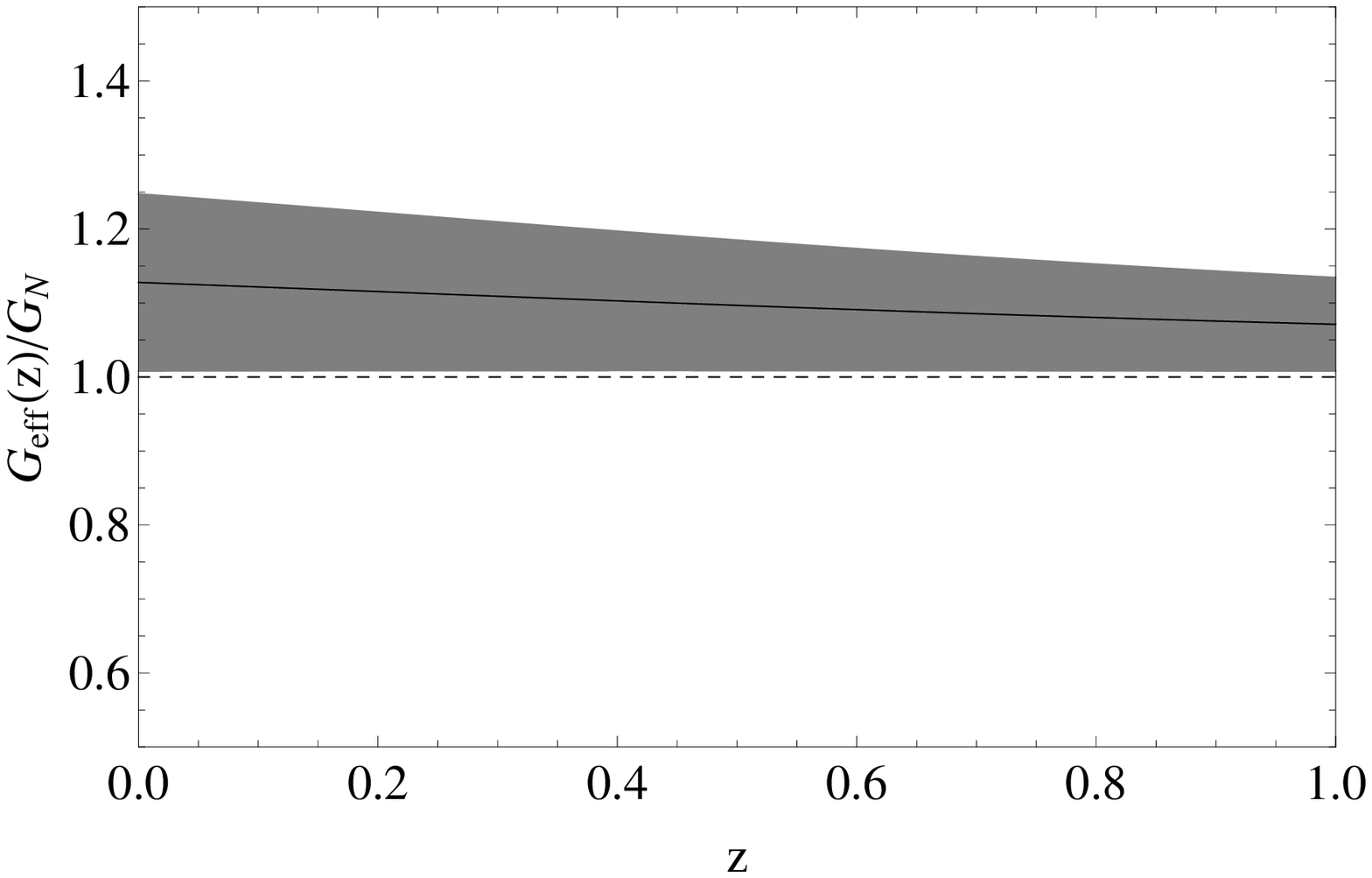}}}
\caption{The evolution of $\Geff(z)$ for the $f(R)$ (left) and $f(T)$ (right) models. The dashed line corresponds to GR, while the solid black line corresponds to the best fit and the gray regions to the $1\sigma$ error region. \label{geffplot}}
\end{figure*}

Finally, in Fig. \ref{geffplot} we show the evolution of $\Geff(z)$ for the $f(R)$ (left) and $f(T)$ (right) models. The dashed line corresponds to GR, while the solid black line  corresponds to the best fit and the gray regions to the $1\sigma$ error region. Not surprisingly, even though these models share the same expansion history, their evolution of Newton's constant $\Geff$ can be quite different in principle. However, as seen in this plot as well, it is not possible to differentiate between these theories and GR with the current data.

\subsection{Forecasts}
We will now try to estimate the necessary improvement in future data so that we can discriminate these degenerate theories from $\Lambda$CDM. This is similar in spirit as Ref. \cite{Nesseris:2011pc}, where the number of new data points needed to detect a potentially  evolving $\Geff$ was estimated. Here however, we will follow a different strategy and assume that the number of points will remain more or less the same, but the quality of the data will improve, i.e. they will have smaller errors. We model this improvement in the data with the error improvement factor
\be
s\equiv \frac{\sigma_{i,old}}{\sigma_{i,new}},
\label{impfac}
\ee
or in other words the factor $s$ expresses how much better the new data have to be compared to the current ones. \footnote{For details regarding the growth rate data used in the current analysis, see Ref.~\cite{Nesseris:2013jea}.}

\begin{figure*}[t!]
\centering
\vspace{0cm}\rotatebox{0}{\vspace{0cm}\hspace{0cm}\resizebox{0.48\textwidth}{!}{\includegraphics{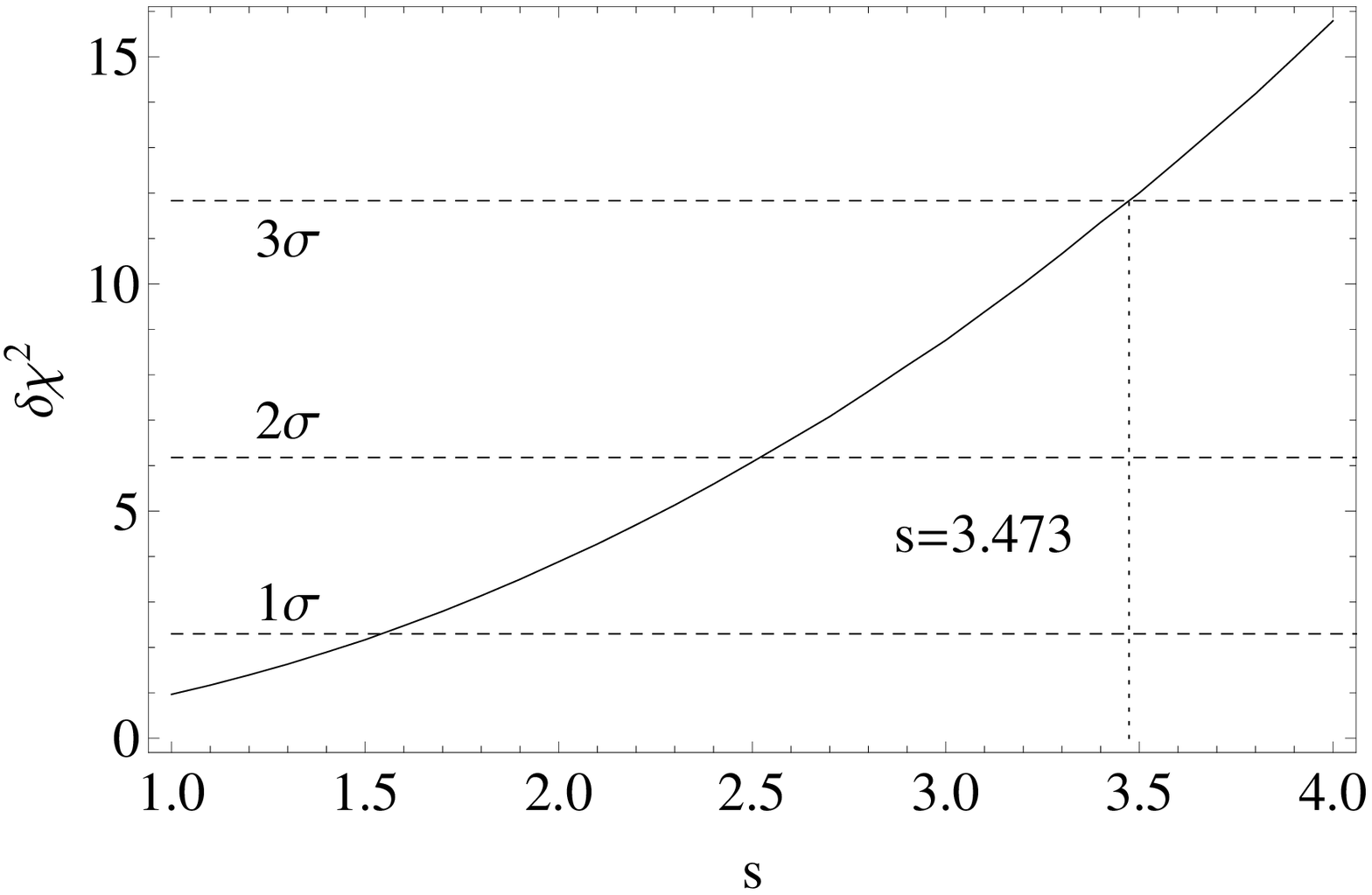}}}
\vspace{0cm}\rotatebox{0}{\vspace{0cm}\hspace{0cm}\resizebox{0.48\textwidth}{!}{\includegraphics{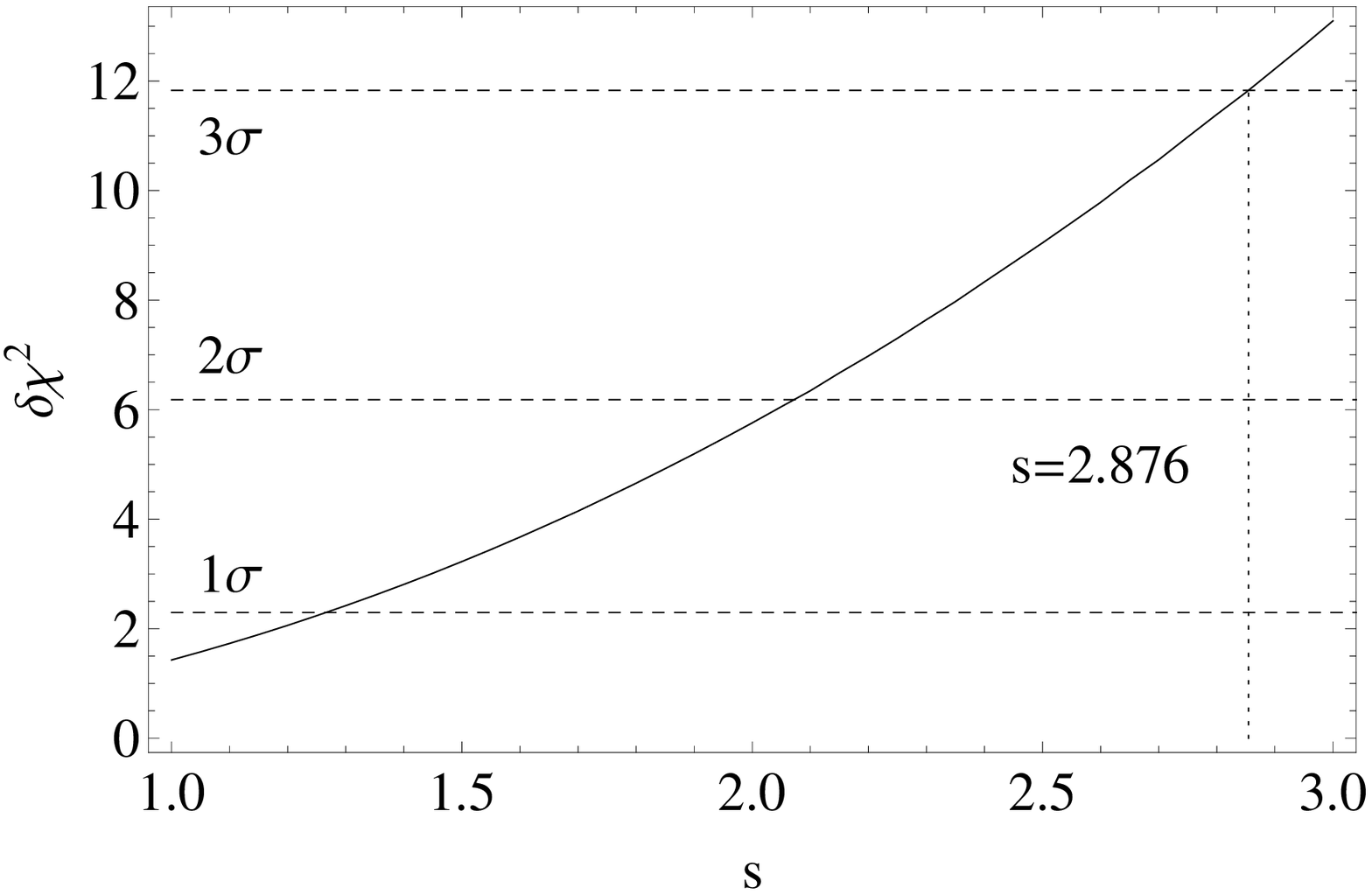}}}
\caption{The difference $\delta \chi^2(s)=\chi^2(s,\alpha=0)-\chi^2_{min}(s)$ as a function of the error improvement factor $s$ for the $f(R)$ (left) and $f(T)$ (right) models. The horizontal lines correspond to the 1, 2 and $3\sigma$ levels, while the vertical dashed lines indicate the value of $s$ for which a $3\sigma$ difference is found. In both cases, $s\sim3$ which implies that a threefold improvement in the growth rate data is needed in order to discriminate these degenerate models and \lcdmb.\label{errorimprovement}}
\end{figure*}

In Fig. \ref{errorimprovement} we show the difference $\delta\chi^2(s)=\chi^2(s,\alpha=0)-\chi^2_{min}(s)$ as a function of the error improvement factor $s$ for the $f(R)$ (left) and $f(T)$ (right) models. In this context the $\delta\chi^2$ expresses the statistical significance of the difference between the best fit and GR. As it can be seen, in order to have a minimum of a $3\sigma$ difference an improvement of $s\sim 3.5$ is needed for the $f(R)$ model, while an improvement of $s\sim 2.9$ is needed for the $f(T)$ model. This is also depicted in Fig. \ref{contours} by the dashed contours, which are estimated in a manner such that the \lcdm best fit sits exactly at the edge of the $3\sigma$ contour. So, in order to differentiate these models from GR and \lcdm at the $3\sigma$ level, future growth rate data will have to improve by at least a threefold, assuming always a constant number of points. To be more concise, a threefold improvement means that the error bars should decrease from $\sigma \in [0.04, 0.18]$, see Table 1 of Ref. \cite{Basilakos:2013nfa}, to $\sigma \in [0.013, 0.06]$. Such a drastic improvement will be possible in the next decade or so with a survey like Euclid; see for example Ref. \cite{Amendola:2012ys}.

\section{Conclusions}
We have investigated a wide range of different modified gravity models, with up to one free parameter, based on the widely popular $f(R)$, $f(T)$ and $f(G)$ theories, both at the background and at the perturbation levels. All of these models were specifically constructed so that they are degenerate, i.e. identical, with respect to the \lcdm model at the background level and as a result no geometric test, like the SnIa, the CMB or the BAO, can discriminate them. Also, it is easy to see that at the perturbations level their evolution can be quite different, e.g. by having a time-dependent Newton's constant $\Geff$ or a quite different $\gamma(z)$ evolution. Therefore, the goal of this analysis was to test whether the present cosmological data are sufficiently good so as to discriminate these degenerate models from \lcdm and if not, what kind of improvements will be necessary in future surveys.

In order to break the degeneracy we used the growth rate data, a dynamic probe that tracks the evolution of the matter density perturbations and in principle can discriminate between the different competing theories. However, we found that with the present data it is impossible to differentiate these degenerate models from \lcdmb, since as can be seen in Fig. \ref{contours}, the best fit in the cases of the $f(R)$ and $f(T)$ models sits well within the $1\sigma$ region. This is due to the fact that at the background level both models mimic exactly the \lcdm model, thus providing an excellent fit for the SnIa, CMB and BAO.

Regarding the $f(G)$ model, as mentioned in Secs. \ref{fG} and \ref{resultscomp}, it suffers from instabilities in the UV but also in the evolution of the matter density perturbations unless it is exactly identical to the \lcdmb. To be more specific, a critical parameter of the model is $\mu$, given by Eq. (\ref{mueq1}), which characterizes the deviation of the $f(G)$ model from GR and the \lcdm model. However, this parameter $\mu$ is also linked to the cosmological scales $k$ at which the instabilities appear; see Eq. (\ref{mueq2}). For small values of $\mu$, the scales can be pushed beyond what is observable from surveys, but as shown in this analysis, even in this case the allowed parameter space is very small and the best fit was found to be virtually indistinguishable from the \lcdmb.

We also determined the required improvement in future growth rate data in order to discriminate these theories from $\Lambda$CDM at the $3\sigma$ level. Assuming a constant number of points, we found that at least a threefold improvement is necessary, something which will be possible with a survey like Euclid. The error bars of the growth rate data used in the paper are in the range $\sigma \in [0.04, 0.18]$, see Table 1 of Ref. \cite{Basilakos:2013nfa} and the redshift $z$ ranges from $0.02$ to $0.8$. Therefore, a threefold improvement would mean that the error bars should decrease to $\sigma \in [0.013, 0.06]$. However, as shown in the Euclid Theory Working Group report Ref. \cite{Amendola:2012ys}, the forecasts for the survey indicate that it will be possible to measure the growth rate $f$ to within $1\%-2.5\%$ accuracy in each bin for $z=[0.5,2.1]$, with bin size $dz=0.1$, and the error on $f$ is projected to be in the range $\sigma=[0.009,0.023]$.\footnote{See Table 1.4, pg 89 of Ref. \cite{Amendola:2012ys} for the forecasts.}

For Euclid this improvement will come by reconstructing the clustering of galaxies up to $z=2$ and the pattern of light distortion from weak lensing up to $z=3$, so the requirement for a threefold decrease in the errors or equivalently $\sigma \in [0.013, 0.06]$ seems a realistic expectation. However, it should be stressed that the most interesting region will be $z=[0.8-1.5]$, a fact that has already been stressed in Ref. \cite{Nesseris:2011pc} where it was shown that due to the presence of a sweet spot in the growth-rate at $z\sim1.7$, having more points at redshift higher than $z>1.5$ does not help in discriminating modified gravity from the \lcdm model, so the redshift range of interest is  $z=[0.8-1.5]$. Finally, it should be mentioned that in practice Euclid or the Dark Energy Survey will have not only growth rate data of much better quality, but also a higher number of data points, which along with the current ones will definitely provide even stronger constraints on modified gravity theories.

\section*{Acknowledgements}
The author would like to thank J.~G.~Bellido,  A.~ De La Cruz Dombriz and S.~Tsujikawa for useful discussions. S.N. acknowledges financial support from the Madrid Regional Government (CAM) under the program HEPHACOS S2009/ESP-1473-02, from MICINN under  Grant No. AYA2009-13936-C06-06 and Consolider-Ingenio 2010 PAU (CSD2007-00060), as well as from the European Union Marie Curie Initial Training Network UNILHC PITN-GA-2009-237920. S.N.  also acknowledges the support of the Spanish MINECO's ``Centro de Excelencia Severo Ochoa" Programme under Grant No. SEV-2012-0249.

\appendix
\section{DERIVATION OF EQ.~(\ref{fTanalsoldz}) \label{analsol}}
In order to solve equation (\ref{ode}) in the case of the $f(T)$ model when $\Geff$ is given by Eq.~(\ref{GeffT1}) and the cosmic expansion is given by the \lcdm model of Eq.~(\ref{Ezlcdm}), we found it convenient to change variables from the scale factor $a$ to $E(z)\equiv H(z)/H_0$. The first step is to solve Eq.~(\ref{Ezlcdm}) for $a$ and if we only keep the real solution we have
\be
a(E)=\frac{\sqrt[3]{\omms}}{\sqrt[3]{E^2-\delta \om}},
\ee
where as mentioned earlier $E(z)\equiv H(z)/H_0$ and $\delta \om=1-\omms$. Then, the first and second derivatives of $H(a)$ are
\ba
H'(a)/H_0&=& -\frac{3 \left(E^2-\delta \om\right)^{4/3}}{2 E \sqrt[3]{\omms}},\\
H''(a)/H_0&=&\frac{3 \left(E^2-\delta \om\right)^{5/3} \left(3 \delta \om+5 E^2\right)}{4 E^3 \omms^{2/3}}.
\ea
Changing variables from $a$ to $E(a)$, using the value of $\Geff(a)$ from Eq.~(\ref{GeffT1}) and using the above equations, results in a differential equation for the growth factor $\delta=\delta(E(a))$ given by
\be
\delta''(E)= \frac{2 E \left((E-\tilde{\alpha} ) \delta'(E)-\delta(E)\right)}{3 \left(\delta \om-E^2\right) (E-\tilde{\alpha} )},\label{newode}
\ee
where we have set $\tilde{\alpha}\equiv\frac{\alpha}{2\sqrt{6}}$. In order to find the proper initial conditions we remember that for a matter dominated universe ($\omms=1$) in GR ($\tilde{\alpha}=0$) we have $E(a)=a^{-3/2}$ and we expect the general solution to be
\ba
\delta_{g, GR}(a)&=&k_1\;a^{-3/2}+k_2\;a\nn \\
&=&k_1\;E(a)+k_2\;E(a)^{-2/3}.
\ea
So, in this formalism we identify the initial conditions for the growing mode as $\delta_g(E)\sim E(a)^{-2/3}$.

Now, equation (\ref{newode}) can easily be solved, e.g. with the help of Mathematica, and the general solution is given by Eq.~(\ref{fTanalsoldz}) as mentioned in the text. In order to find the growing mode we demand that at early times, i.e. when $E\gg1$, the growth factor behaves as $\delta_g(E)\sim E(a)^{-2/3}$. Doing so, we find that the constants $K_1$ and $K_2$ have to be given by Eqs.~(\ref{K1}) and (\ref{K2}) for the growing mode.

\section{THE PERTURBATION EQUATIONS \label{pert}}
Here we have gathered some of the equations used in Sec. \ref{fG}. It can be shown that $\delta_m$ satisfies the evolution equation \cite{DeFelice:2009rw}:

\begin{equation}
\label{delmeq}
\ddot{\delta}_m+C_1 \dot{\delta}_m+C_2 \delta_m=
r \ddot{\Phi}_2+(d_8-r d_4)\dot{\Phi}_2\,,
\end{equation}
where $C_1 \equiv r d_1-d_5$,
$C_2 \equiv r d_2-d_6$, and
\begin{equation}
r \equiv \frac{M_B^2}{M_A^2}, \quad
M_A^2 \equiv d_3+c_2^2\frac{k^2}{a^2},\quad
M_B^2 \equiv d_9+d_7\frac{k^2}{a^2},
\end{equation}

\begin{eqnarray}
\hspace*{-2.7em}& &c_2^2=\frac{1+\Omega_m+2x+4\mu (1+2x)}{1+4\mu}, \\
\hspace*{-2.7em}& & d_1=\frac{\mu [1+\Omega_m+2x+4\mu (1+2x)]}{H(1+4\mu)}, \\
\hspace*{-2.7em}& & d_2=\frac{\Omega_m [1+3\Omega_m+4x+4\mu (1+4x)]}{4(1+4\mu)}, \\
\hspace*{-2.7em}& & d_3=H^2 \{ 4x^2+4x+x'-2\mu [4(1-3x^2-2x-x') \nonumber \\
\hspace*{-2.7em}& &~~~+3\Omega_m(1+x)+8\mu (2-2x^2-x')] \}/
[2\mu (1+4\mu)],
 \label{d3} \\
\hspace*{-2.7em}& & d_4=-\frac{3H[1+\Omega_m+2x+4\mu (1+2x)]}{1+4\mu},
\end{eqnarray}

\begin{eqnarray}
\hspace*{-2.7em}& & d_5= -\frac{2H (1-2x \mu+2\mu)}{1+4\mu},\quad
d_6=\frac{3H^2 \Omega_m (1+x)}{1+4\mu},\\
\hspace*{-2.7em}& & d_7=\frac{4H^2 (1+x)}{1+4\mu},\quad
d_8=-\frac{12 H^3 (1+x)}{1+4\mu},\\
\hspace*{-2.7em} & & d_9=\frac{3H^4[4x^2+4x+x'-4\mu (1-3x^2-2x-x')]}
{\mu (1+4\mu)},
\label{d9}
\end{eqnarray}

where $x \equiv \dot{H}/H^2$, $x' \equiv \dot{x}/H$ and $\Omega_m \equiv
8\pi G\rho_m/(3H^2)$. Also we have defined the important parameters:

\begin{eqnarray}
\label{mudef}
\xi &\equiv& f_{\GB},\\
\mu &\equiv& H \dot{\xi}=H\dot{\GB}f_{,\GB\GB} \nonumber \\
&=& 72H^6f_{\GB \GB} \left[ (1+w_{\rm eff})(1+3w_{\rm eff})-\frac{w_{\rm eff}'}{2}
\right].
\end{eqnarray}

\end{document}